\documentclass[9pt,conference]{IEEEtran}
\IEEEoverridecommandlockouts
\usepackage{multicol,multirow}
\usepackage{amsmath,amsfonts,graphicx}
\usepackage{amssymb,textcomp,mathtools}
\usepackage{bm,upgreek,algorithm,hyperref}
\usepackage{multirow,booktabs,hhline,array}
\usepackage{cite,url,makecell,setspace}
\usepackage{xcolor}
\pdfoutput=1

\title{Time-Frequency-Based Attention Cache Memory Model for Real-Time Speech Separation\thanks{This work was supported by the National Key Research and Development Program of China (No. 2021ZD0200301) and the National Natural Science
Foundation of China (No. U2341228). \\ $^*$Guo Chen and Kai Li contribute equally to the article. $^\dagger$Corresponding author}}
\author{\IEEEauthorblockN{Guo Chen$^{1,*}$, Kai~Li$^{1,*}$, Runxuan Yang$^{1}$, Xiaolin Hu$^{1,2,\dagger}$\\}
\IEEEauthorblockA{$^1$Department of Computer Science and Technology, BNRist, \\
THBI, IDG/McGovern Institute for Brain Research, Tsinghua University, Beijing, China \\
$^2$Chinese Institute for Brain Research (CIBR), Beijing, China\\
cg22@mails.tsinghua.edu.cn, tsinghua.kaili@gmail.com, yangrx20@mails.tsinghua.edu.cn, xlhu@tsinghua.edu.cn}} 

\begin{document}
\maketitle

\begin{abstract}
Existing causal speech separation models often underperform compared to non-causal models due to difficulties in retaining historical information. To address this, we propose the Time-Frequency Attention Cache Memory (TFACM) model, which effectively captures spatio-temporal relationships through an attention mechanism and cache memory (CM) for historical information storage. In TFACM, an LSTM layer captures frequency-relative positions, while causal modeling is applied to the time dimension using local and global representations. The CM module stores past information, and the causal attention refinement (CAR) module further enhances time-based feature representations for finer granularity. Experimental results showed that TFACM achieveed comparable performance to the SOTA TF-GridNet-Causal model, with significantly lower complexity and fewer trainable parameters. For more details, visit the project page: \url{https://cslikai.cn/TFACM/}.
\end{abstract}
\begin{IEEEkeywords}
Causal model, Speech Separation, Time-frequency Domain 
\end{IEEEkeywords}

\section{Introduction}
\label{sec:introduction}
Speech separation involves extracting individual audio from mixed audio signals, enabling clearer and more intelligible communication in complex environments.
Recently, with the rapid development of deep learning techniques, 
the performance of speech separation was improved significantly \cite{luo2018tasnet,luo2020dual,li2022efficient,hu2021speech,li22d_interspeech,li2024iianet}. Currently, the extensively studied task in speech separation is offline speech separation, which processes recorded audio in batches, allowing it to utilize both past and future information from the entire recording. 
Many of the state-of-the-art (SOTA) speech separation methods are developed for offline tasks, and they are called non-causal models\cite{wang2023tf,subakan2021attention,chen2020dual, li2024spmamba}.

In contrast, real-time tasks, such as conference transcription \cite{yoshioka2019advances}, need to be done in real-time, and future information cannot be utilized. Models that handle real-time tasks are called causal models.
Causal models \cite{wu2020end,li2022skim,della2024resource} require the system to acquire and process information only from the current and preceding time frames. Despite advances, causal models still lag behind non-causal counterparts in performance \cite{luo2019conv}. Efforts to narrow this gap typically involve extending the look-ahead window \cite{schroter2022low} or implementing joint-training strategies \cite{li22d_interspeech,li2023design}. These methods use additional tricks to enable the model to directly or indirectly utilize future information. Recent studies have explored integrating historical hidden layer states to enhance causal models by extracting past audio feature representations, thereby improving efficiency \cite{li2022skim,li2023predictive}. Although existing causal models are capable of storing historical information, their performance in separation remains suboptimal.

TF-GridNet \cite{wang2023tf} is a SOTA method for non-causal time-frequency domain separation, and it is known for its robustness in handling complex real-world audio environments. The model simultaneously processes temporal and frequency components, enabling it to capture intricate patterns in audio data. The structure of TF-GridNet has the potential to be modified for causal tasks. 
However, due to the large number of parameters in TF-GridNet, directly improving it to a causal version would result in slower inference and be unsuitable for real-time scenarios, while directly compressing the parameters of TF-GridNet could cause performance degradation
(see Section \ref{sec:sota} for details). This raises the question of how to efficiently retain historical information in a time-frequency-based model, such as TF-GridNet, when adapting it for causal processing.

We introduce a novel causal speech separation model called TFACM, designed to capture spatio-temporal relationships within an attention-based dual-path framework. TFACM utilizes LSTM to model relative positions along the frequency dimension, followed by causal modeling along the time dimension through segmentation into local and global representations. A Cache Memory (CM) module is incorporated to store historical information during global modeling. Additionally, a Causal Attention Refinement (CAR) module fine-tunes attention weights using past alignment information, leading to more precise feature representations. Experimental results on public datasets demonstrated that TFACM achieved comparable separation performance to the SOTA method, TF-GridNet-causal. Furthermore, TFACM exhibits only 8.8\% of the parameters and 20.4\% of the computational complexity of TF-GridNet-causal, significantly reducing resource requirements.


\section{TF-based Attention Cache Memory Model}
\label{sec:model}
\subsection{Model Overview}

\begin{figure*}[h]
	\small
	\centering
	\includegraphics[width=2.0\columnwidth]{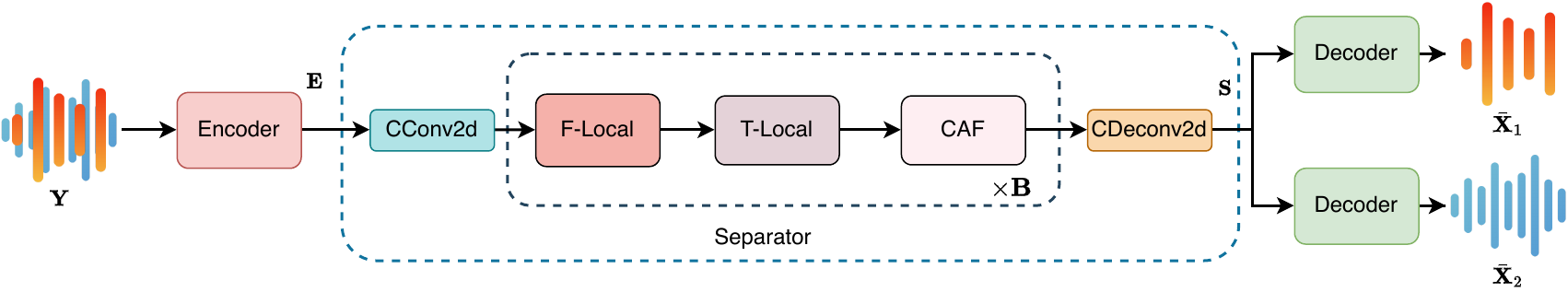}
	\caption{The pipeline of TFACM.}
	\label{fig:separation-pipeline}
 \vspace{-15pt}
\end{figure*}

The proposed model is built upon the architecture of TF-GridNet \cite{wang2023tf}, comprising three main components: the encoder, the separator which encodes both time and frequency domains with LSTMs and attention module, and the decoder, as illustrated in Fig. \ref{fig:separation-pipeline}. The input mixture signal which contains multiple speaker signals $\mathbf{X}_c\in \mathbb{R}^{1\times L}$ and noise $\mathbf{N}\in \mathbb{R}^{1\times L}$, is denoted as  $\mathbf{Y}=\sum_{c=1}^{C}\mathbf{X}_c+\mathbf{N}, \mathbf{Y}\in \mathbb{R}^{1\times L}$, where $C$ denotes the number of speakers, and $L$ denotes the original signal length. The encoder initially applies an STFT to convert the input signal $\mathbf{X}$ into a time-frequency feature representation $\mathbf{E}\in \mathbb{R}^{2\times F \times T}$, where ``2" denotes the feature dimension, $T$ denotes the number of time steps, and $F$ denotes the number of frequency bands.
The separator processes the input features by leveraging both temporal and frequency dimensions to generate the processed embedding sequence for each output channel. 
The decoder is composed of an iSTFT layer, which reconstructs the processed embeddings back into time-domain signals $\bar{\mathbf{X}}_c\in \mathbb{R}^{1\times L}$.

\begin{figure*}[h]
	\small
	\centering
	\includegraphics[width=2.0\columnwidth]{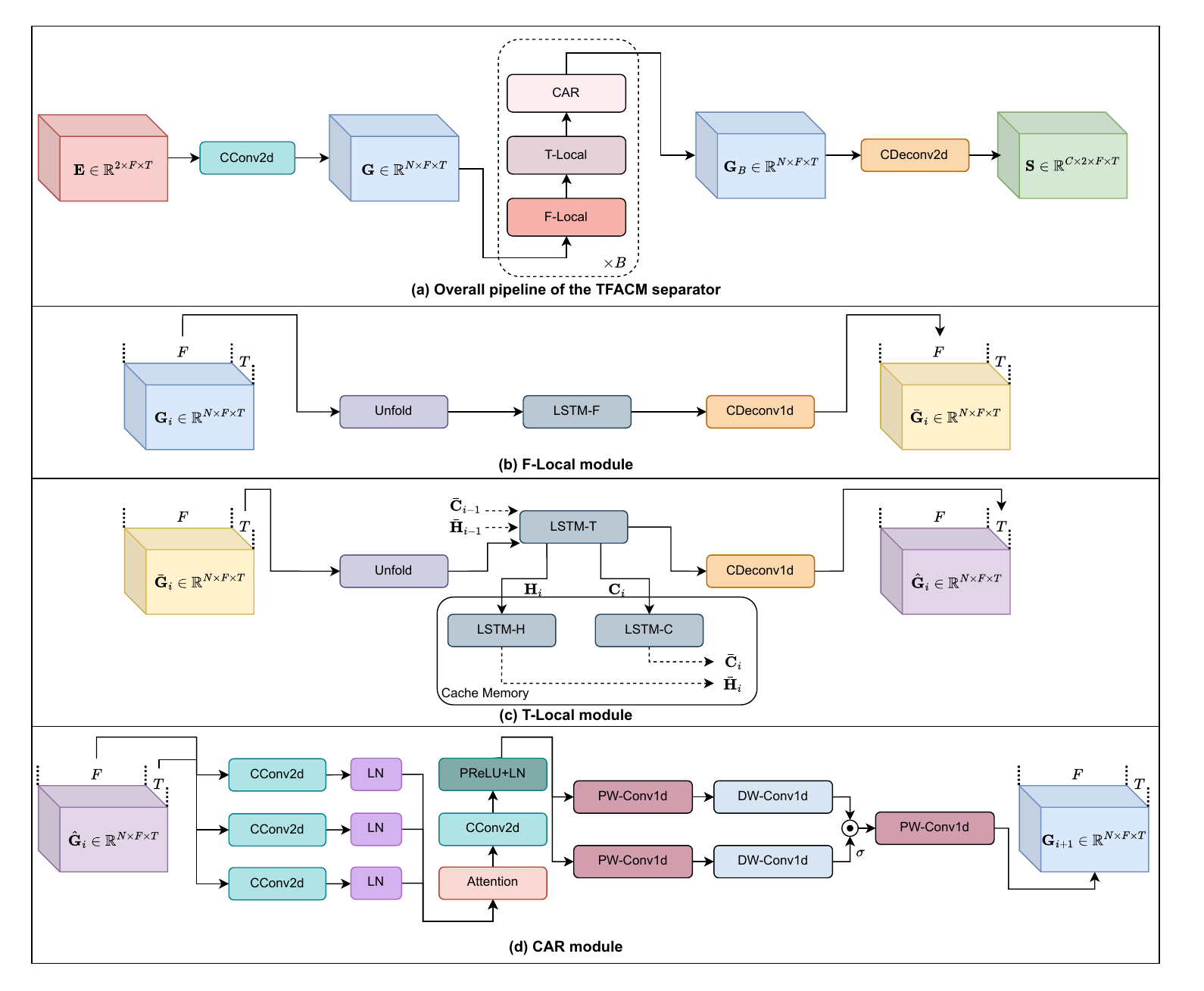}
	\caption{The pipeline of the TFACM separator. PW/DW-Conv represents the point-wise/depth-wise convolutional layer.}
	\label{fig:tfacm}
 \vspace{-18pt}
\end{figure*}

\subsection{A Naive Extension to TF-GridNet}\label{sec:causal-set-tfg}

In this section, we first introduce a naive extension of TF-GridNet, named TF-GridNet-Causal, which is the framework used in TFACM. To transform TF-GridNet into a causal model, we need three key components: a convolutional layer, an LSTM layer, and an attention module.

\textbf{Convolutional Layers}: TF-GridNet comprises a series of convolutional layers and transposed convolutional layers, which are non-causal in the original TF-GridNet.
Convolutional layers are non-causal because they use future frame information when generating the output at time step $T$. To make the model causal, these non-causal layers are replaced with causal ones, achieved by adding a zero vector to the latter half of the kernel or masking the weights of the latter half. The causal 1D/2D convolutional and transpose convolutional layers are referred to as CConv1d/CConv2d and CDeconv1d/CDeconv2d, respectively.

\textbf{LSTM Layers}: In TF-GridNet, LSTM layers are primarily employed for temporal and frequency modeling. In the original TF-GridNet, the use of bidirectional LSTMs allows each time step to access information from future frames
However, in constructing the causal model, we replaced the bidirectional LSTM with a unidirectional LSTM to ensure that the model relied solely on past information.

\textbf{Attention Module}: In a non-causal setting, the self-attention module \cite{vaswani2017attention} allows each frame to access information from all frames, including future ones. In contrast, the causal model uses a diagonal attention mask, restricting each frame’s access to only past information, ensuring inference relies solely on prior frames.

\subsection{Separator of TFACM}
\label{sec:separator}

As shown in Fig. \ref{fig:tfacm}, the TFACM separator follows a two-stage pipeline to separately model frequency and time dimensions. The input $\mathbf{E}$ is first processed by a CConv2d layer and a Layer Normalization (LN) layer \cite{ba2016layer}, transforming the time-frequency features into a high-dimensional representation $\mathbf{G} \in \mathbb{R}^{N \times F \times T}$, where $N$ is the number of feature channels. Next, $\mathbf{G}$ is sequentially processed through $B$ TFACM separator blocks. Each block contains an F-Local module, a T-Local module, and a causal attention refinement (CAR) module, iteratively updating $\mathbf{G}_i$ to $\mathbf{G}_{i+1}$. The input for the first block, $\mathbf{G}_0$, is the output from the initial transformation. These modules capture both local and global time-frequency information to enhance separation performance. Finally, a CDeconv2d layer with $2 \times C$ output channels, a $3 \times 3$ kernel, and ReLU activation generates the predicted real and imaginary representations $\mathbf{S} \in \mathbb{R}^{C \times 2 \times F \times T}$. Each component of the TFACM separator is explained in detail in the following parts.

\textbf{F-Local Module}: The design of the F-Local module is similar to the intra-frame spectral module in TF-GridNet, but we replace the non-causal components with their causal counterparts. Specifically, we first apply an unfold operation to segment the frequency dimension of $\mathbf{G}_i\in \mathbb{R}^{N\times F \times T}$ into sub-band features $\mathbf{G}^F_i\in \mathbb{R}^{W_1\times N\times L_F \times T}$, using a fixed width $W_1$ and stride $S_1$, where $L_F=\frac{F-W_1}{S_1}+1$.
Next, we apply LN across the channel dimension, followed by an unidirectional LSTM layer, called LSTM-F, to obtain $\bar{\mathbf{G}}^F_i\in \mathbb{R}^{H\times L_F \times T}$, where $H$ denotes the hidden channels of features. In LSTM-F, the feature dimension is $W_1\times N$ and the sequence length is $L_F$. Finally, a CDeconv1d layer is used to integrate features across the sub-bands, generating $\bar{\mathbf{G}}_i\in \mathbb{R}^{N\times F \times T}$.

\textbf{T-Local Module}: To mitigate historical information loss along the temporal dimension $T$, we propose a Cache Memory (CM) module, serving as a shared repeater across the $B$ TFACM separator blocks. It leverages LSTM hidden states $(\mathbf{H}, \mathbf{C})$ to facilitate cross-module historical information interaction, as illustrated in Fig. \ref{fig:tfacm}(c). The module processes input $\bar{\mathbf{G}}_i$ by segmenting the time dimension into sub-band features $\bar{\mathbf{G}}^T_i \in \mathbb{R}^{W_2 \times N \times F \times L}$ using a fixed window width $W_2$ and stride $S_2$, where $L = \frac{T - W_2}{S_2} + 1$. These features are then passed through the LSTM layer (LSTM-T), which also receives previous hidden states $(\bar{\mathbf{H}}_{i-1}, \bar{\mathbf{C}}_{i-1}) \in \mathbb{R}^{2 \times H \times F \times L}$ from the CM module. The output is $\hat{\mathbf{G}}^T_i \in \mathbb{R}^{W_2 \times H \times F \times L}$ along with updated hidden states $(\mathbf{H}_i, \mathbf{C}_i)$, where $(\mathbf{H}_i, \mathbf{C}_i) = ([\mathbf{h}_{i,1}, \dots, \mathbf{h}_{i,L}], [\mathbf{c}_{i,1}, \dots, \mathbf{c}_{i,L}])$.In LSTM-T, the feature dimension is $N$ and the sequence length is $W_2$, equivalent to running $L$ independent LSTMs with identical parameters, each processing a segment of length $W_2$. This design allows efficient modeling of long-term dependencies across temporal segments.
This sequence $(\mathbf{H}_i, \mathbf{C}_i)$ is then re-encoded using the CM module and passed on to the next T-Local block. The sequence $(\mathbf{H}_i, \mathbf{C}_i)$ is processed to ensure causality, facilitating information transfer between different TFACM separator blocks. For each hidden state $(\mathbf{h}, \mathbf{c})$ of the $i$-th TFACM separator block, the cache model encodes these states as follows:
\begin{equation}
(\mathbf{H}'_{i}, \mathbf{C}'_{i}) = \text{CacheMemory}(\mathbf{H}_i, \mathbf{C}_i),
\end{equation}
where CacheMemory represents encoding of $\mathbf{H}_i$ and $\mathbf{C}_i$ by two LSTMs, called LSTM-H and LSTM-C, respectively. This encoding process, by reprocessing the hidden states, aids in preserving and enhancing temporal information.

Since the hidden layer parameters of each LSTM-T encapsulate the complete information for their corresponding time segments, in order to ensure that $\mathbf{H}'_{i}$ and $\mathbf{C}'_{i}$ do not experience information leakage when transmitted to the ($i+1$)-th block, special misalignment process is required.
Specifically, this process is as follows:
\begin{equation}
(\bar{\mathbf{h}}_{i,l},\bar{\mathbf{c}}_{i,l}) = \left\{
\begin{aligned}
& (\mathbf{0}, \mathbf{0}) \quad  &l=1 \\
& (\mathbf{h}'_{i,l-1},\mathbf{c}'_{i,l-1}) \quad &l > 1
\end{aligned}
\right..
\end{equation}
Thus, $(\bar{\mathbf{H}}_{i}, \bar{\mathbf{C}}_{i})$ is returned to the LSTM-T of the $(i + 1)$-th block  as a comprehensive input that summarizes the global information. Finally, the output features $\hat{\mathbf{G}}^T_i$ are used as the input of the CDeconv1d layer to obtain $\hat{\mathbf{G}}_{i}\in \mathbb{R}^{N\times F \times T}$.

\textbf{CAR Module}: After separately processing the frequency and time dimensions, an attention mechanism enhances the time-frequency features. High-dimensional features $Q$, $K$, and $V \in \mathbb{R}^{N \times F \times T}$ are extracted using a CConv2d layer with PReLU and LN. To maintain causality, a diagonal mask is applied in the multi-head attention module to block future frame information. Following the attention module, a gated convolution method is used, comprising two parallel paths: one path applies pointwise and depthwise convolution, followed by an activation function, while the other path applies the same convolutions without activation. These paths are merged and passed through a final pointwise convolution layer, with the gate-selecting features based on the time-frame memory. 

\begin{table*}[h]
\footnotesize
\centering
\caption{Comparison with different causal speech separation methods across three datasets, along with parameters and MACs. The MACs and inference time were measured on the GPU for 1 second of audio at a sample rate of 8 kHz. Bolded numbers indicate the best results among all models, and underlined numbers represent the second-best results.}
\begin{tabular}{cccccccccc}
\toprule
\multirow{2}{*}{Methods} & \multicolumn{2}{c}{WHAM!}     & \multicolumn{2}{c}{WHAMR!} & \multicolumn{2}{c}{LibriMix}  & \multirow{2}{*}{Params (M)} & \multirow{2}{*}{MACs (G/s)} & \multirow{2}{*}{Time (ms)} \\  \cmidrule(r){2-3} \cmidrule(r){4-5} \cmidrule(r){6-7}
                         & SDRi    & SI-SNRi & SDRi & SI-SNRi & SDRi    & SI-SNRi &                               &          &                 \\ \midrule
DPRNN\cite{luo2020dual}                    & 11.0          & 10.6          & 7.8        & 7.2           & 9.6           & 9.0           & 1.3                           & 11.2      &   \textbf{33.42}                \\
SkiM\cite{li2022skim}                     & 10.8          & 10.4          & 7.7        & 7.0           & 9.9           & 9.2           & 3.6                           & \underline{11.1}     &     \underline{39.01}               \\
ReSepFormer\cite{della2024resource}              & 11.7      & 11.3      & 7.6        & 6.9           & 4.1           & 3.3           & 8.0                           & \textbf{6.3}          & 47.46               \\ 
TF-GridNet-Causal \cite{wang2023tf} & \textbf{13.8}          & \textbf{13.5}          & \textbf{9.9}        & \textbf{9.2}           & \underline{11.5}           & \underline{10.8}           & 11.4                           & 179.2   &        84.59              \\ 
\midrule
TFACM (Small)                 & 12.0 & 10.9 & \underline{9.6}          & \underline{9.0}             & 10.2 & 9.3 & \textbf{0.5}                           & 19.4       & 45.17                 \\ 
TFACM (Large)                 & \underline{13.3} & \underline{13.0} & 9.6          & 8.9             & \textbf{11.7} & \textbf{11.0} & \underline{1.0}                           & 36.5       & 57.00                 \\ 
\bottomrule
\end{tabular}
    
    \label{tab:causal-performance-all}
    \vspace{-15pt}
\end{table*}

\section{Experiment configurations}
\label{sec:config}
\subsection{Dataset}

To enhance the realism and challenge of our experiments, we selected complex datasets, including WHAM!\cite{wichern2019wham}, WHAMR!\cite{maciejewski2020whamr}, and LibriMix\cite{cosentino2020librimix}. The WHAM! dataset augments WSJ0-2mix by introducing background noise from real environments. The WHAMR! dataset further extends WSJ0-2Mix by adding both noise and reverberation effects. The LibriMix dataset, based on the LibriSpeech corpus, is a widely-used benchmark in speech separation research. All models were trained and tested using an 8 kHz sampling rate.

\subsection{Model configurations}

The parameter settings for TFACM were kept consistent across training and testing on different datasets. We used an STFT window size of 8 ms, a hop size of 1 ms, and a Hann window. The feature dimension $N$ was set to 128 for TFACM (Large) and 64 for TFACM (Small). The number of TFACM blocks was $B = 3$ for the Large model and $B = 2$ for the Small model. Each LSTM layer within the TFACM blocks had 64 hidden units. In the CAR module, the number of attention heads was set to 2 for the Large model and 4 for the Small model.

\subsection{Training and evaluation}
During training, we used the negative Signal-to-Noise Ratio (SNR) between the model output and the clean target as the objective function, employing the uPIT strategy \cite{yu2017permutation} to handle label permutation in speech separation. Training was conducted on 8 Nvidia 3090 GPUs using the Adam optimizer with an initial learning rate of 0.001. Gradient clipping with a maximum norm of 5 was applied. If no improvement was observed in 10 consecutive epochs, the learning rate was halved. Training stopped after 15 epochs without validation improvement. For evaluation, we used SI-SNRi \cite{le2019sdr} and SDRi \cite{vincent2006performance} to assess separation performance. Model complexity was measured by the number of parameters and MACs\footnote{\url{https://github.com/sovrasov/flops-counter.pytorch}}. The TFACM source code is available at \url{https://anonymous.4open.science/r/TFACM-Code/}.

\section{Results}
\label{sec:result}

\subsection{Comparison with SOTA methods}
\label{sec:sota}

We conducted a quantitative analysis of TFACM’s performance in comparison with existing causal speech separation methods.
The experimental results are presented in Table \ref{tab:causal-performance-all}. The Table \ref{tab:causal-performance-all} showed that TFACM (Large) demonstrated performance metrics comparable to those of TF-GridNet-Causal, and exceeded those of other models. 
Although TF-GridNet-Causal slightly outperforms TFACM (Large) in separation performance, its large number of parameters and high computational complexity poses challenges for deployment in low-latency causal applications. In contrast, TFACM uses only 8.8\% of the parameters and incurs only 20.4\% of the computational cost of TF-GridNet-Causal, significantly reducing model complexity.

To further optimize computational efficiency, we propose TFACM (Small), a lightweight version designed for low-complexity scenarios. Compared to models with as many as 8 million parameters, such as ReSepFormer, TFACM (Small) demonstrates substantial advantages in architectural efficiency while surpassing 
those models
in performance. Notably, TFACM (Small) achieves an inference time of 45.17 milliseconds, which falls within acceptable limits for practical applications. This indicates that the model effectively balances performance and computational complexity, making it suitable for real-world deployment.

\subsection{Comparison on different audio input length}

\begin{figure}[h]
	\small
	\centering
	\includegraphics[width=0.8\columnwidth]{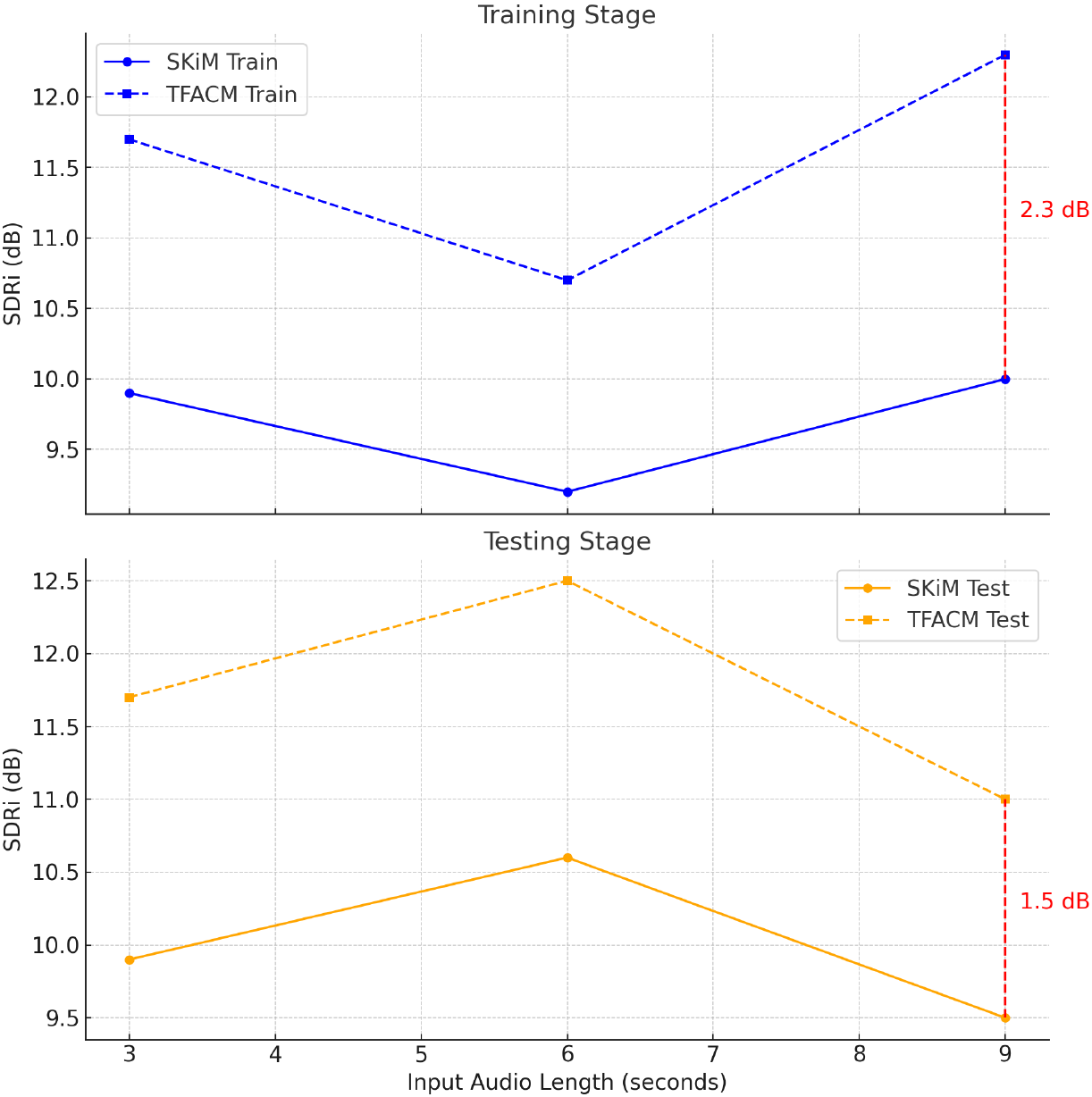}
	\caption{Comparison with longer audio input at the training stage and the testing stage.}
	\label{fig:comparison_longer_audio}
 \vspace{-20pt}
\end{figure}

We conducted experiments on the max subset\footnote{\url{https://github.com/JorisCos/LibriMix}} of the Libri2Mix dataset \cite{cosentino2020librimix}, which includes audio inputs ranging from 6 to 10 seconds, to evaluate our module’s ability to handle longer audio sequences. In the first experiment, we extended the training audio length from the original 3 seconds to 6 and 9 seconds for comparison. Using SkiM \cite{li2022skim} as the baseline, results in Table \ref{fig:comparison_longer_audio} show that TFACM maintained strong performance with increasing input length. In the second experiment, we fixed the training length at 3s but tested with longer inputs segmented into overlapping shorter segments. The results again demonstrated the TFACM’s robustness and ability to generalize across different input lengths.

\subsection{Ablation Studies on the CM and CAR Modules}

We evaluated the effectiveness of the CM and CAR modules on the WHAM! dataset \cite{wichern2019wham}. For the CM module, we replaced LSTM-H and LSTM-C with an unaligned identity mapping, effectively passing the state of the current block directly to the next TFACM block. For the CAR module, we experimented with completely removing the gated convolution after the attention module. The results of these ablation studies are presented in Table \ref{tab:ablation}. We found that removing the CM module led to a decrease in overall performance, likely due to the model losing its ability to process internal information, which significantly weakened its handling of temporal information. Furthermore, removing the CAR module also resulted in a significant performance drop, mainly because the CAR module provides coarse-grained global time-frequency perception by integrating separately processed frequency and temporal features. These findings further underscore the necessity and effectiveness of incorporating the CM and CAR modules within the TFACM model.

\begin{table}[h]
    \centering
    \vspace{-5pt}
    \caption{Ablation Studies of TFACM on WHAM!}
    \vspace{-5pt}
    \begin{tabular}{cccc}
    \toprule
        CM module & CAR module & SDRi & SI-SNRi \\ \midrule
        \checkmark & \checkmark & \textbf{13.3} & \textbf{13.0} \\ \midrule
        $\times$&  \checkmark& 11.8 & 11.4 \\
        \checkmark &  $\times$& 11.8 & 11.5 \\ 
        $\times$ & $\times$ & 11.2 & 10.7 \\ \bottomrule
    \end{tabular}

    \label{tab:ablation}
    \vspace{-15pt}
\end{table}

\section{Conclusion}
\label{sec:conclusion}
We introduce a causal speech separation method using the TFACM model and provide a detailed performance analysis. This model incorporates a cache memory module and a causal attention refinement module to utilize past information while maintaining causality effectively. Experimental results demonstrated that TFACM surpassed existing causal models, especially in complex reverberant environments, underscoring its potential for real-world use. This study highlights the significance of integrating historical hidden states and refining causal attention in improving causal speech separation models, offering valuable insights for future system designs.

\bibliographystyle{IEEEtran}
\bibliography{refs}

\end{document}